\documentclass[letterpaper,twocolumn,english,pre]{revtex4}
\usepackage[T1]{fontenc}
\usepackage[latin9]{inputenc}
\usepackage{xcolor}
\usepackage{pdfcolmk}
\usepackage{soul}
\usepackage{amsmath}
\usepackage{graphicx}

\providecolor{lyxadded}{rgb}{0,0,1}
\providecolor{lyxdeleted}{rgb}{1,0,0}

\usepackage{babel}

\begin{document}

\title{Anomalous uniform domain in a twisted nematic cell constructed from
micropatterned surfaces}

\author{T. J. Atherton}

\email{timothy.atherton@case.edu}

\affiliation{Case Western Reserve University, 10900 Euclid Avenue, Cleveland,
Ohio, USA. 44106}

\author{J. R. Sambles}

\affiliation{Electromagnetic Materials Group, School of Physics, University of
Exeter, Stocker Road, Exeter, England. EX4 4QL}

\author{J. P. Bramble}

\author{J. R. Henderson}

\author{S. D. Evans}

\affiliation{Molecular and Nanoscale Physics Group, School of Physics and Astronomy,
The University of Leeds, Woodhouse Lane, Leeds, LS2 9JT}
\begin{abstract}
We have discovered an optically uniform type of domain that occurs
in Twisted Nematic (TN) cells that are constructed from substrates
chemically patterned with stripes via microcontact printing of Self-Assembled
Monolayers (SAM); such domains do not occur in TN cells constructed
from uniform substrates. In such a cell, the azimuthal anchoring at
the substrates is due to the elastic anisotropy of the liquid crystal
rather than the conventional rubbing mechanism. A model is presented
that predicts the relative stability of the twisted and anomalous
states as a function of the material and design parameters.
\end{abstract}
\maketitle

\section*{Introduction}

One of the most commercially important liquid crystal devices is the
twisted nematic (TN) cell \citep{schadt:127}, which comprises a nematic
liquid crystal film sandwiched between two substrates that have been
coated with a rubbed polymer film to promote planar alignment with
orthogonal azimuthal easy axes. The local axis of rotational symmetry
of the alignment, known as the director, must rotate azimuthally through
the depth of the cell in order to satisfy the boundary conditions
at both substrates. The total rotation may be either $+\pi/2$ or
$-\pi/2$ radians; these two degenerate configurations are referred
to arbitrarily as \emph{twist }and \emph{reverse-twist} and regions
of the cell in each configuration are separated by domain walls as
they are topologically distinct. To lift the degeneracy, a polymer
that promotes an easy axis at some small angle to the plane of the
substrate is used \citep{0305-4470-11-7-030}. 

Micropatterned surfaces consist of micron-scale regions that promote
different easy axes. Away from the distorted surface region, the liquid
crystal adopts a bulk orientation that depends on the pattern; by
appropriate patterning, it is possible to design a surface to promote
an arbitrary \emph{effective }easy axis. In this spirit, Lee and Clark
experimentally showed that a surface micropatterned with alternate
planar degenerate and homeotropic stripes will align the liquid crystal
azimuthally even though the planar regions are azimuthally degenerate
\citep{Baek-woonLee03302001}. One of the present authors later showed
analytically \citep{Atherton:2006p31} that in contrast to the usual
mechanism, it is the elastic anisotropy that determines the preferred
alignment direction so as to minimize the elastic energy. If the twist
elastic constant $K_{2}$ is less than the splay $K_{1}$ and bend
$K_{3}$ elastic constants, as is the case for most nematic materials,
the azimuthal easy axis lies along the length of the stripes. The
polar easy axis is determined by the relative width of the planar
and homeotropic stripes. 

In the present work, we use a striped surface made by microcontact
printing of self assembled monolayers (SAMs) of $\omega$-functionalised
alkanethiols \citep{Cheng2000,Gupta:1997p2000,Bramble:2008p2382}
on gold to demonstrate that this elastic-anisotropy mechanism is sufficiently
strong to permit the construction of a twisted nematic cell. Since
the homeotropic-planar mark-space ratio is 1:1, the cells have a high
tilt angle ($\sim45^{\circ}$) and exhibit a novel uniform domain
forbidden in a conventional TN cell.

\section*{Experiment}

\begin{figure*}
\begin{centering}
\includegraphics{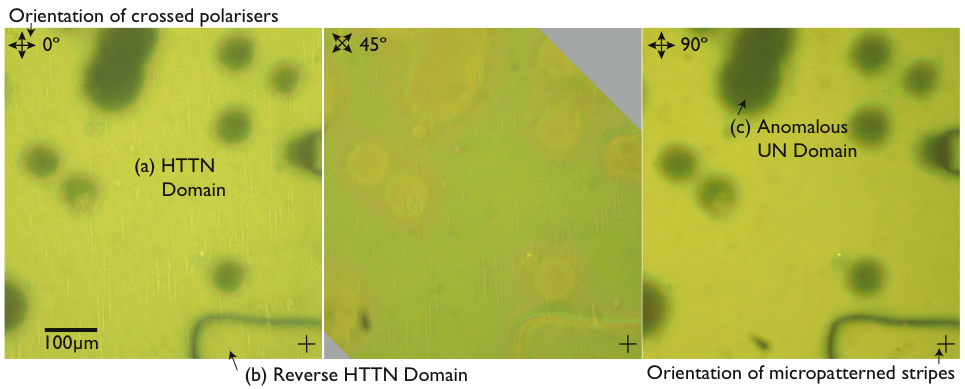}
\par\end{centering}

\caption{\label{cap:Polarising}Polarising microscope images of a twisted nematic
cell constructed from two striped micropatterned substrates. The cell
thickness $d=20\mu m$ and the period of patterning $\lambda=2\mu m$.
The arrows on the upper left of each image show the alignment of crossed
polarizers; the crosses on the bottom right show the orientation of
the stripes. Various regions are observed: (a) twisted domain, (b)
reverse twisted domain and (c) planar domain.}

\end{figure*}

The micropatterned twisted nematic cell was constructed as follows
in cleanroom conditions: Glass microscope slides were used as substrates
and were thoroughly cleaned with ultrasonic agitation in detergent
(Decon 90), ultra pure water and acetone. The slides were rinsed in
water and immersed in pirahna etch (70\% sulphuric acid, 30\% hydrogen
peroxide) for 20 mins. The slides were rinsed and dried in nitrogen
before being loaded into a thermal evaporator (Edwards Auto 306).
A 3nm layer of Cr was evaporated first, to aid the adhesion of a subsequent
30nm layer of Au. The slides were rotated during evaporation to ensure
uniform coverage and to avoid any anisotropy in the roughness of the
gold surface which can lead to liquid crystal alignment effects \citep{Gupta:1996p1999}.

PDMS (Sylgard 184, Dow Corning) was formed in a ratio of 9 parts elastomer
to 1 part curing agent. The mixture was well stirred, degassed under
vacuum and poured into a mould containing a Si wafer with 1$\mu$m
etched stripes. Prior to this, the wafer was immersed in a 1$\%$
solution of perflurotriethoxysilane (Fluorochem Ltd.) in dichloromethane
for 12 hours. This silane monolayer ensures that the cured PDMS can
be removed from the etched wafer without breaking \citep{Michel2001}.
The PDMS was cured in an oven at 60$^{\circ}$C for 12 hours, carefully
cut up into stamps and finally rinsed in ethanol to remove any short
chain PDMS material \citep{Kraus2005}. 

The PDMS stamps were inked on the printing surface with a 7mM solution
in ethanol of perfluorinated alkanethiol ($(CF_{3})-(CF_{2})_{17}-(CH_{2})_{11}-SH$),
which is known to promote homeotropic alignment in 6CB\citep{Alkhairalla1999}.
The inking solution was allowed to diffuse into the PDMS for approximately
3 mins and then the printed surface was dried with nitrogen. The stamp
was then placed on a gold substrate for 3 mins. Good quality printing
occurs when the stamp is in conformal contact with the substrate \citep{Bietsch2000},
which can be observed during the printing process. The printed surface
was then immersed for 12 hours in a 3mM solution in ethanol of mercapto-undecanoic
acid ($COOH-(CH_{2})_{11}-SH$), which is known to promote planar
degenerate alignment in 6CB\citep{Alkhairalla1999}. The micropatterned
surfaces were then rinsed in clean ethanol and water and then dried.

A cell was constructed from two patterned substrates and 23$\mu$m
PET spacers by first placing the liquid crystal 6CB (Merck) on one
surface, heating into the isotropic phase and then placing the second
patterned substrate on top with the stripes orthogonal to the first
substrate. This procedure was adopted as LC cells which contain perfluorinated
alkanethiol SAMs are difficult to fill with liquid crystal by capillary
action. The completed cell was then cooled to room temperature.

When viewed between crossed polarizers under the microscope, a number
of domains become apparent (fig. \ref{cap:Polarising}): the majority
of the cell {[}fig. \ref{cap:Polarising}(a)] appears bright and remains
bright when the cell is rotated azimuthally with the polarisers fixed;
other regions {[}fig. \ref{cap:Polarising}(b)] behave in a similar
manner, but are separated from the first by dark lines; the final
class of regions observed in fig. \ref{cap:Polarising}(c) appear
dark when the striped surfaces are aligned with the polarizers, but
become bright when the cell is rotated with the maximum intensity
at $45^{\circ}$. The latter regions are not separated from the former
by a distinct line, rather the intensity of transmitted light varies
smoothly over a distance of a few microns from that of the surrounding
region (which is of the first or second type) to that of the enclosed
region (of the third type).

\begin{figure}
\begin{centering}
\includegraphics{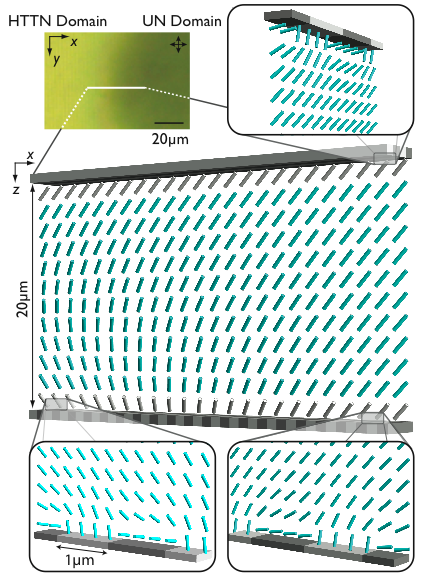}
\par\end{centering}

\caption{\label{fig:Schematic}Schematic of the director configuration across
a transition region between the HTTN and UN configurations. The micropatterning
causes rapid variation of the director near the substrates (see insets)
which enforces an average effective orientation as a boundary condition
on the bulk configuration (main figure).}

\end{figure}

The first and second regions are twist and reverse twist domains typical
of a twisted nematic cell and so we shall refer to them as High Tilt
Twisted-Nematic (HTTN) states: the director rotates azimuthally by
$\pi/2$ from the bottom substrate to the top substrate. These configurations
act as polarisation converters and thus permit transmission independent
of azimuthal angle. The optical behaviour of the uniform third region\textendash{}denoted
hereafter the Uniform Nematic (UN) state\textendash{}is identical
to that of a planar cell and so the director in this region must be
confined to a plane defined by the surface normal and a vector that
points in the direction of one of the sets of stripes. In fig. \ref{fig:Schematic},
a possible director profile is shown for the HTTN and UN configurations
and an experimentally observed transition region. The UN domains are
of particular interest because they are not observed in a conventional
twisted nematic cell. No regions are observed that are dark irrespective
of orientation with respect to the polarizers (the behaviour of a
homeotropic cell).

\section*{Model \& Discussion}

We now construct a model for the observed behaviour by evaluating
the energy of the two observed configurations. Let us define a coordinate
system local to the surface where the $x$-axis lies along the wavevector
of the stripes, the $y$-axis lies along the length of the stripes
and the $z$-axis lies perpendicular to the substrate. The director
may be parameterized\begin{equation}
n=(\cos\theta\cos\phi,\cos\theta\sin\phi,\sin\theta),\label{eq:director}\end{equation}
then the Frank-Oseen elastic energy density\begin{equation}
f_{el}=\frac{1}{2}K_{1}(\mathbf{\nabla\cdot\mathbf{\mathbf{\hat{n}}}})^{2}+\frac{1}{2}K_{2}(\mathbf{\hat{n}}\cdot\nabla\times\mathbf{\hat{n}})^{2}+\frac{1}{2}K_{3}|\mathbf{\hat{n}}\times\nabla\times\mathbf{\hat{n}}|^{2},\label{eq:frankenergydensity}\end{equation}
within the approximation $K_{1}=K_{3}\neq K_{2}$ has, as was shown
in \citep{Atherton:2006p31}, the form\begin{equation}
f_{el}(\kappa)=\frac{K_{1}}{2}(\kappa\theta_{x}^{2}+\theta_{z}^{2})\label{eq:freeenergy}\end{equation}
where $\kappa=[1-(1-\tau)\sin^{2}\phi]$ in which $\tau=K_{2}/K_{1}$
and where the subscripts indicate derivates taken with respect to
these coordinates. The Euler-Lagrange equation for the configuration
$\theta(x,z)$ is a scaled version of Laplace's equation,\begin{equation}
\kappa\theta_{xx}+\theta_{zz}=0\label{eq:thetaeulerlagrange}\end{equation}
with the series solution\begin{align}
\theta(x,z)=\theta_{0}+2 & \sum_{n=1}^{\infty}\exp(-2n\pi\sqrt{\kappa}z/\lambda)\times\nonumber \\
 & \ \ \times\left[p_{n}\sin(2n\pi x/\lambda)+q_{n}\cos(2n\pi x/\lambda)\right]\label{eq:fouriersoln}\end{align}
where the coefficients $\theta_{0}$, $p_{n}$ and $q_{n}$ are determined
from the boundary condition. By substituting \eqref{eq:fouriersoln}
into \eqref{eq:freeenergy} and integrating, a general expression
for the elastic energy per unit area may be obtained\begin{align}
F_{el}(\kappa) & =\frac{1}{\lambda}\int_{0}^{\infty}\int_{0}^{1}\int_{0}^{\lambda}f_{el}(\text{\ensuremath{\kappa}})\ \text{d}x\text{d}y\text{d}z\nonumber \\
 & =\frac{K_{1}}{\lambda}2\pi\sqrt{\kappa}\sum_{n=1}^{\infty}n(p_{n}^{2}+q_{n}^{2}).\label{eq:elasticenergyfouriergeneral}\end{align}

\begin{flushleft}
For a harmonic anchoring potential \begin{equation}
W_{\theta}(\theta-\theta_{e})^{2}/2\label{eq:harmonicanchoringpotential}\end{equation}
where\begin{gather}
\theta_{e}=\begin{cases}
\frac{\pi}{2} & 0<x\leq a\\
0 & a<x\leq1\end{cases}\label{eq:BC}\end{gather}
the natural boundary condition is\begin{equation}
\left[\theta_{z}+2\frac{1}{L_{\theta}}(\theta-\theta_{e})\right]_{z=0}=0\label{eq:boundary}\end{equation}
where $L_{\theta}=K_{1}/(W_{\theta}\lambda)$ is a characteristic
penetration depth of the surface treatment. Since the planar SAMs
promote azimuthally degenerate anchoring, there is no azimuthal anchoring
term to be considered. Substituting \eqref{eq:boundary} into \eqref{eq:fouriersoln}
and exploiting the orthogonality of the $\sin$ and $\cos$ functions,
the coefficients\begin{align}
\theta_{0} & =\frac{\pi a}{2},\ p_{n}=\frac{\sin^{2}(na\pi)}{2n(1+2n\pi L_{\theta}\sqrt{\kappa})},\nonumber \\
q_{n} & =\frac{\sin(2na\pi)}{4n(1+2n\pi L_{\theta}\sqrt{\kappa})}\label{eq:weakcoeff}\end{align}
are obtained. The bulk elastic energy per unit area may be evaluated
by substituting \eqref{eq:weakcoeff} into \eqref{eq:elasticenergyfouriergeneral}\begin{equation}
F_{el}(\kappa)=\frac{K_{1}}{\lambda}\pi\sqrt{\kappa}\sum_{n=1}^{\infty}\frac{\sin(na\pi)^{2}}{2n(1+2n\pi L_{\theta}\sqrt{\kappa})^{2}}.\label{eq:elastic energy}\end{equation}
The surface anchoring energy per unit area is obtained by substituting
\eqref{eq:fouriersoln} evaluated at $z=0$ and \eqref{eq:BC} into
\eqref{eq:harmonicanchoringpotential} and integrating\begin{eqnarray}
F_{surf}(\kappa) & = & \frac{W_{\theta}}{2}\frac{1}{\lambda}\int_{0}^{1}\int_{0}^{\lambda}[\theta(x,0)-\theta_{e}(x)]^{2}\ \text{d}x\text{d}y\nonumber \\
 & = & \frac{W_{\theta}}{2}\left[\frac{\pi^{2}a(1-a)}{4}-\vphantom{-\sum_{n=1}^{\infty}\frac{\left(1+4n\pi L_{\theta}\sqrt{\kappa}\right)\sin^{2}(na\pi)}{2m^{2}\left(1+2n\pi L_{\theta}\sqrt{\kappa}\right)^{2}}}\right.\nonumber \\
 &  & \ \left.-\sum_{n=1}^{\infty}\frac{\left(1+4n\pi L_{\theta}\sqrt{\kappa}\right)\sin^{2}(na\pi)}{2m^{2}\left(1+2n\pi L_{\theta}\sqrt{\kappa}\right)^{2}}\right].\label{eq:fsurfaceenergy}\end{eqnarray}

\par\end{flushleft}

It is now possible to estimate the respective energies of the HTTN
state and the UN states using the above expressions. In the UN state
the director is oriented at some constant azimuthal angle $\psi$
with respect to the wavevector of the patterning on the lower substrate;
there is no bulk elastic deformation and the energy is thus the energy
of the distorted surface regions. Since there are two perpendicularly
patterned substrates some care must be taken in identifying $\psi$
with $\phi$, which was defined above with reference to the pattern
on one substrate. The total energy of the UN state per unit area,
$F_{UN}$, is \begin{eqnarray}
F_{UN} & = & F_{el}(\kappa_{lower})+F_{surf}(\kappa_{lower})+\nonumber \\
 &  & +F_{el}(\kappa_{upper})+F_{surf}(\kappa_{upper})\label{eq:UNenergy}\end{eqnarray}
where $\kappa_{lower}=[1-(1-\tau)\sin^{2}\psi]$ is associated with
the lower substrate and $\kappa_{upper}=[1-(1-\tau)\cos^{2}\psi]$
is associated with the upper substrate. This energy has a minimum
value when $\psi=0$ or $\psi=\pi/2$ i.e. when the director is aligned
along the length of one or the other of the patterns, which is consistent
with the experimentally observed configuration above.

In the HTTN state, the azimuthal orientation is fixed at the surfaces
by the elastic anisotropy mechanism such that the component of the
director in the plane of the substrate lies along the length of the
stripes, and the azimuthal orientation rotates by $\pi/2$ through
the cell thickness. Since the period of the micropatterning $\lambda\ll d$
the cell thickness, in the middle of the cell the director depends
on the $z$-coordinate only. The characteristic penetration length
of the distorted surface regions is of order $\lambda$ and so the
small azimuthal rotation of the director over this length ought to
leave the energy due to a distorted surface region unchanged from
the situation above where $\phi$ is a constant $\phi=\pi/2$. The
surface contribution to the HTTN state is therefore\begin{equation}
F_{HTTNsurface}=2F_{el}(\tau)+2F_{surf}(\tau).\label{eq:FHTTNsurface}\end{equation}

\begin{figure}
\begin{centering}
\includegraphics{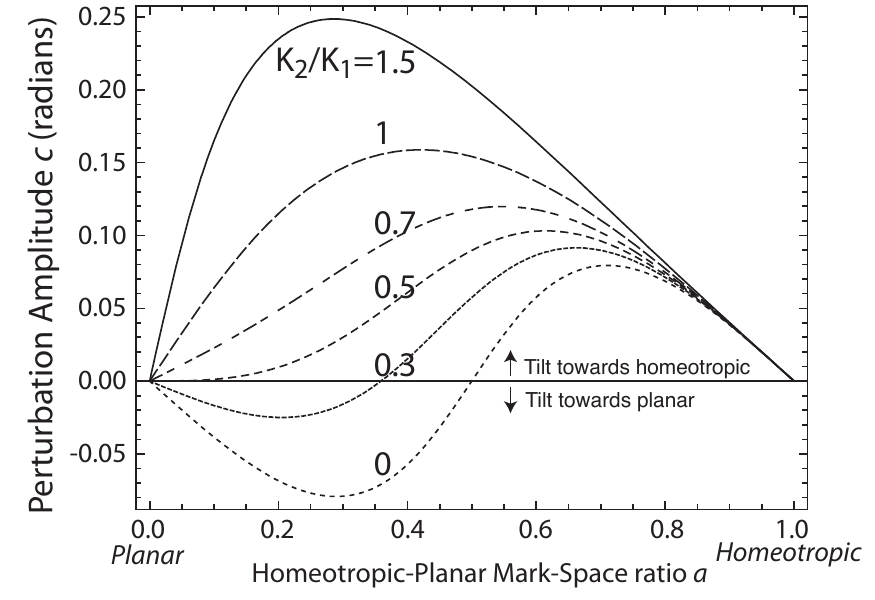}
\par\end{centering}

\caption{\label{fig:pertmagnitude}Magnitude of perturbation of the tilt angle
as a function of the homeotropic-planar mark-space ratio $a$ plotted
for several different ratios of the elastic constants $K_{2}/K_{1}$.}

\end{figure}

It is also necessary to include the contribution from the azimuthal
rotation of the director in the bulk. In contrast to the conventional
TN cell, the polar angle is not constant throughout the cell; the
director can reduce the magnitude of the twist deformation by tilting
towards homeotropic in the centre of the cell at the expense of a
slight splay or bend deformation. From recent simulation of the HTTN
state \citep{Atherton:2007p2495}, a suitable ansatz for the bulk
director configuration is\begin{equation}
\theta(z)=\frac{\pi a}{2}+c\sin(\pi z/d),\ \ \phi(z)=\frac{\pi z}{2d}\label{eq:ansatz1dhttn}\end{equation}
where the amplitude of the perturbation $c$, the magnitude of the
tilt perturbation, is a free parameter that must be varied so as to
minimize the total free energy. Routine substitution of \eqref{eq:ansatz1dhttn}
into the Frank energy \eqref{eq:frankenergydensity} yields after
integration an energy per unit area\begin{align}
F_{HTTNbulk}= & \int_{0}^{d}f_{el}(\theta(z),\phi(z))\ \text{d}z\nonumber \\
= & \frac{K_{1}\pi^{2}}{64d}\left[1+16c^{2}+3\tau+4\tau J_{0}(2c)\cos(a\pi)-\right.\nonumber \\
 & \ -(1-\tau)J_{0}(4c)\cos(2a\pi)-\nonumber \\
 & \ -4\tau\sin(a\pi)H_{0}(2c)+\nonumber \\
 & \left.\ \ +(1-\tau)\sin(2a\pi)H_{0}(4c)\right],\label{eq:Fhttnbulk}\end{align}
where $J_{0}$ is a Bessel function of the first kind and $H_{0}$
is the Struve function. The energy $F_{HTTNbulk}$ must be minimized
numerically with respect to $c$. Typical perturbation amplitudes
are plotted in fig. \ref{fig:pertmagnitude} as a function of the
homeotropic-planar pattern mark-space ratio $a$ which is assumed
to be the same on both substrates; above a certain threshhold value
of $K_{2}/K_{1}$, the tilt perturbation is always toward the homeotropic
orientation, i.e. $c$ is always positive; below this value the tilt
perturbation is towards planar for low $a$ and towards homeotropic
for high $a$. The energy of the HTTN state $F_{HTTN}$ is then \begin{equation}
F_{HTTN}=F_{HTTNsurface}+\min(F_{HTTNbulk})\label{eq:Fhttnbulksum}\end{equation}

\begin{figure}
\begin{centering}
\includegraphics{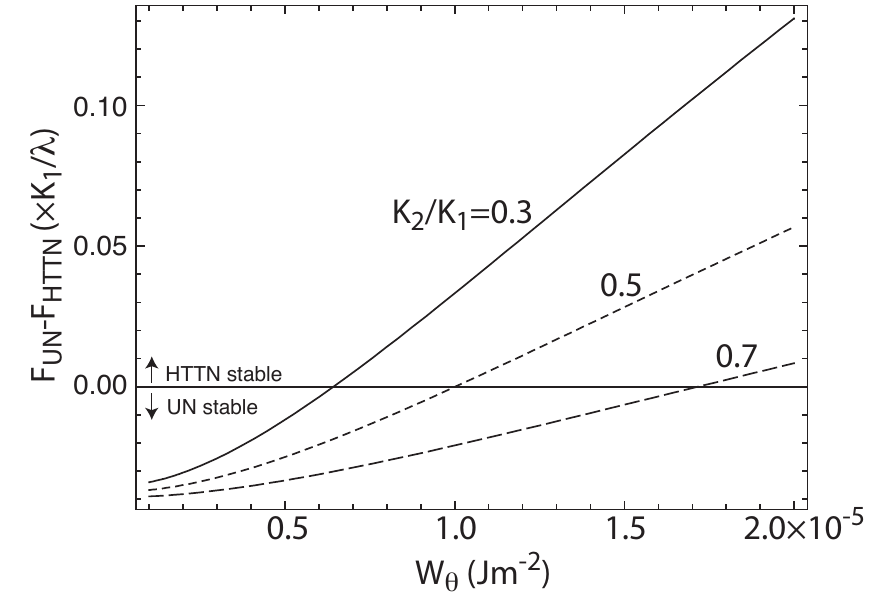}
\par\end{centering}

\caption{\label{fig:EnergyDifference}Energy difference between the Uniform
Nematic (UN) and High Tilt Twisted Nematic (HTTN) states.}

\end{figure}

From the expressions for $F_{HTTN}$ and $F_{UN}$, the difference
between the energies of the two states was calculated for a variety
of polar anchoring strengths $W_{\theta}$ and elastic constant ratios
$K_{2}/K_{1}$ (fig. \ref{fig:EnergyDifference}), using the design
parameters for the experimental pattern, i.e. $\lambda=2\mu\text{m}$
and $a=1/2$ and the absolute magnitude of $K_{1}=1\times10^{-11}\text{N}$
which is comparable with experimental measurements for 6CB \citep{karat77,Madhusudhana:1982p2506};
the critical anchoring energy below which the UN state has lower energy
than the HTTN state is about $1\times10^{-5}\text{Jm}^{-2}$. 

\section*{Conclusion}

A twisted nematic cell has been constructed using substrates micropatterned
with SAMs by microcontact printing. Characteristic twisted and reverse
twisted states have been observed in the cell, however these are stabilized
not by azimuthal anchoring as in a conventional TN device, but by
a recently elucidated elastic anisosotropy mechanism due to rapid
variation of the polar easy axis. This mechanism promotes rather weak
azimuthal anchoring which, if the polar anchoring energy is sufficiently
weak, may be of comparable energy to the bulk twist; facilitating
a transition from the twist or reverse-twist states to a uniform state
that is not facile in conventional TN cells due to the much higher
azimuthal anchoring energy. We found the critical polar anchoring
energy for such a transition to be $W_{\theta}\sim1\times10^{-5}\text{Jm}^{-2}$. 

A further feature of the micropatterned TN cell is that the pretilt
angle is much larger ($\sim45^{\circ}$) than in a conventional TN
display and readily controllable by adjusting the design parameters
of the patterning. An interesting phenomenon has been elucidated that
should be exhibited by any high pretilt TN device regardless of the
anchoring mechanism: there is a spontaneous tendency for the director
to tilt towards the planar or homeotropic orientations depending on
the relative magnitude of $K_{1}$ and $K_{2}$.

The SAMs technique has for the first time been demonstrated to permit
the fabrication of device-scale patterns, and the rich phase diagrams
that arise in cells constructed from micropatterned surfaces suggest
the possibility of interesting additional phenomena such as bistability.
\begin{acknowledgments}
The authors would like to thank Sharp Laboratories of Europe for financial
support and N. J. Smith, S. A. Jewell and C. Rosenblatt for helpful
discussions. This work was supported by the Engineering and Physical
Research Council, Grant No. GR/S59826/01.
\end{acknowledgments}

\end{document}